\documentclass[12pt]{article}
\usepackage{amsfonts}
\usepackage{amssymb}
\usepackage{amsmath}
\usepackage{graphics}
\begin{document}
{\LARGE \centerline{Canonical transformations}
\centerline{in}\centerline{three-dimensional phase space}}

\phantom{aaa}

\phantom{aaa}

{\large \centerline{T. Dereli}}%
\centerline{Physics Department, Ko\c{c} University, 80910
Sar{\i}yer-Istanbul, TURKEY}%
\centerline{{\it tdereli@ku.edu.tr}}

{\large \centerline{A. Te\u{g}men\footnote{On sabbatical from
Physics Department,
Ankara University 06100 Ankara TURKEY}}}%
\centerline{Feza G\"{u}rsey Institute, 34684
\c{C}engelk\"{o}y-Istanbul,
TURKEY}%
\centerline{{\it tegmen@science.ankara.edu.tr}}

{\large \centerline{T. Hakio\u{g}lu}}%
\centerline{Physics Department, Bilkent University, 06533 Ankara, TURKEY}%
\centerline{{\it hakioglu@fen.bilkent.edu.tr}}

\begin{abstract}
Canonical transformation in a three-dimensional phase space
endowed with Nambu bracket is discussed in a general framework.
Definition of the canonical transformations is constructed as
based on canonoid transformations. It is shown that generating
functions, transformed Hamilton functions and the transformation
itself for given generating functions can be determined by solving
Pfaffian differential equations corresponding to that quantities.
Types of the generating functions are introduced and all of them
is listed. Infinitesimal canonical transformations are also
discussed. Finally, we show that decomposition of canonical
transformations is also possible in three-dimensional phase space
as in the usual two-dimensional one.
\end{abstract}
\noindent\rule{7in}{0.01in}
\section{Introduction}
In 1973 Y. Nambu proposed a generalization of the usual
Hamiltonian dynamics, in which odd-dimensional phase spaces are
also possible \cite{ref:Nambu}. To his proposal, time evolution of
a dynamical variable $f(x_1,\dots ,x_n)=f(x)$ over an
$n$-dimensional phase space is given by the so-called Nambu
bracket
\begin{eqnarray}\label{Nambu}
\dot{f}=\{f,H_1,\dots ,H_{n-1}\}=\frac{\partial (f,H_1,\dots
,H_{n-1})}{\partial (x_1, \dots ,x_n)},
\end{eqnarray}
where $H_1,\dots ,H_{n-1}$ are the functionally independent
Hamilton functions and the variables $x_1,\dots ,x_n$ stand for
the local coordinates of $\mathbb{R}^n$. The explicit form of the
Nambu bracket (\ref{Nambu}) is given by the expression
\begin{equation}\label{explicit}
\{f_{1},\dots ,f_{n}\}=\frac{\partial (f_{1},\dots
,f_{n})}{\partial (x_1,\dots ,x_n)}= \epsilon_{i_{1}\cdots
i_{n}}\frac{\partial f_1}{\partial x_{i_1}}\cdots \frac{\partial
f_n}{\partial x_{i_n}}.
\end{equation}
(Throughout the text, sum is taken over all repeated indices). The
coordinate-free expression of the Nambu bracket is defined by
means of the $(n-1)$-form $\Gamma =dH_1\wedge \cdots \wedge
dH_{n-1}$, namely
\begin{eqnarray}
^*(df\wedge \Gamma)=\{f,H_1,\dots ,H_{n-1}\},
\end{eqnarray}
where $d$ and $\wedge$ denote the usual exterior derivative and
exterior product respectively, and $^*$ is the Hodge map.

It is well known that canonical transformations (CTs) are a
powerful tool in the usual Hamilton mechanics. They serve three
main purposes: to describe the evolution of a dynamical system, to
show the equivalence of two systems, and mostly to transform a
system of interest into a simpler or known one in different
variables. In this paper we study  CTs in the phase space endowed
with canonical Nambu bracket and we will try to gain a deeper
insight to the subject in a general framework.

The paper is organized as follows: In Sec.2, a precise definition
of CT in three-space is given. Since every CT is a canonoid
transformation it is felt that an explicit definition of the
canonoid transformations should be given. In doing so, the
discussion is kept in its general pattern, i.e., in the time
dependent form. Additionally, direct conditions on a CT
corresponding to the ones in the usual even-dimensional Hamilton
formalism are constructed . Sec.3 is devoted to show how to find
the generating functions (GFs) and the new Hamilton functions.
This section also contains the way to find the CT for given GFs.
It is seen that if one wants to know the GFs, the CT and the new
Hamilton functions, one must solve a Pfaffian differential
equation related with that quantity. Sec.4 stands for the
exemplification of CTs, including the definitions of gauge and
point CTs in three-space. Sec.5 deals with the classification of
CTs. It gives an extensive number of types. All of the possible
eighteen types is listed in six main kinds in Table~\ref{table1}.
As an inevitable part of the presentation, we construct the
infinitesimal transformations (ICTs) in Sec.6. It is shown that
the construction parallels the usual Hamilton formalism such that
ICTs can generate finite CTs. In order to complete the discussion,
in Sec.7 it is shown that a CT in three-space can be decomposed
into a sequence of three minor CTs. This result, in fact, confirms
a well known conjecture saying the same thing in the usual
classical and quantum mechanics.

\section{Definition of Canonical Transformations in Three-Space}
In the definition (\ref{Nambu}), $f$ and Hamilton functions
$H_1,\dots ,H_{n-1}$ do not contain $t$ explicitly. For the sake
of generality we will allow the explicit $t$ dependence. Since,
for the local coordinates $x_1, x_2, x_3$,  the Nambu-Hamilton
equations of motion give
\begin{eqnarray}\label{NH}
\dot{x_i}=\epsilon_{ijk}\frac{\partial H_1}{\partial
x_j}\,\frac{\partial H_2}{\partial x_k},\qquad i,j,k=1,2,3,
\end{eqnarray}
(from now on, all Latin indices will take values $1,2,3$), total
time evolution of a dynamical variable $f(x,t)$ becomes
\begin{eqnarray}
\dot{f}=\{f,H_1,H_2\}+\frac{\partial f}{\partial t}.
\end{eqnarray}
Hence time evolution of the Hamilton functions amounts to the well
known form
\begin{eqnarray}
\dot{H_\alpha}=\frac{dH_\alpha}{dt}=\frac{\partial
H_\alpha}{\partial t},\qquad \alpha =1,2.
\end{eqnarray}

Instead of giving directly the definition of a CT in three-space,
it may be remarkable to give some interesting situations as a
pre-knowledge. First, by using the same terminology developed for
the usual Hamilton formalism in the literature
\cite{ref:Abraham,ref:Jose}, we give the definition of a canonoid
transformation. The main definition of a CT will be based on this
definition.\newline\newline%
{\bf Definition 2.1.} For a dynamical system whose equations of
motion are governed by the pair $(H_1(x,t),H_2(x,t))$, the time
preserving diffeomorphism $\mathbb{R}^3\times
\mathbb{R}\rightarrow \mathbb{R}^3\times \mathbb{R}$ such that
\begin{eqnarray}\label{imap}
(x_i,t)\mapsto (X_i(x,t),t)
\end{eqnarray}
is called a \emph{canonoid }transformation with respect to the
pair $(H_1,H_2)$ if there exist a
pair\newline%
$(K_1(X,t),K_2(X,t))$ satisfying
\begin{eqnarray}\label{covariance}
\dot{X}_i=\epsilon_{ijk}\frac{\partial K_1}{\partial
X_j}\,\frac{\partial K_2}{\partial X_k},
\end{eqnarray}
where $\mathbb{R}^3\times\mathbb{R}$ is the extended phase space
in which $t$ is considered as an additional independent variable.

The invertible transformation (\ref{imap}) (canonoid or not) also
changes the basis of vector fields and differential forms:
\begin{eqnarray}\label{vfield}
\frac{\partial}{\partial x_i}=\frac{\partial X_j}{\partial
x_i}\frac{\partial}{\partial X_j}+\frac{\partial t}{\partial
x_i}\frac{\partial }{\partial t}(=0),\qquad
\frac{\partial}{\partial X_i}=\frac{\partial x_j}{\partial
X_i}\frac{\partial}{\partial x_j}+\frac{\partial t}{\partial
X_i}\frac{\partial }{\partial t}(=0),
\end{eqnarray}
\begin{eqnarray}\label{form}
dx_i=\frac{\partial x_i}{\partial X_j}dX_j+\frac{\partial
x_i}{\partial t}dt(=0),\qquad dX_i=\frac{\partial X_i}{\partial
x_j}dx_j+\frac{\partial X_i}{\partial t}dt.
\end{eqnarray}

In the time independent case, the extended part drops and the map
becomes on $\mathbb{R}^3$ as expected, i.e.,
\begin{eqnarray}
x_i\mapsto X_i(x).
\end{eqnarray}
Note that, such a map considers $t$ in any time dependent function
$f(x,t)$ as a parameter only.

According to Definition 2.1 it is obvious that $K_1$ and $K_2$
serve as Hamilton functions for the new variables and the
transformation (\ref{imap}) preserves the Nambu-Hamilton
equations.

As an example consider Nambu system
\begin{eqnarray}
\dot{x_1}=x_2x_3\;,\;\dot{x_2}=-x_1x_3\;,\;\dot{x_3}=0
\end{eqnarray}
governed by the Hamilton functions
\begin{eqnarray}\label{Hs}
H_1(x)=\frac{1}{2}(x_1^2+x_2^2)\;,\;H_2(x)=\frac{1}{2}x_3^2.
\end{eqnarray}
Let the transformation be
\begin{eqnarray}\label{canonoid}
X_1=x_1\;,\;X_2=x_2\;,\;X_3=x_3^2.
\end{eqnarray}
Now if we choice the new Hamilton functions as
\begin{eqnarray}
K_1(X)=\frac{1}{2}(X_1^2+X_2^2)\;,\;K_2(X)=\frac{2}{3}X_3^{3/2},
\end{eqnarray}
we see that Nambu-Hamilton equations of motion remain covariant.
For a different pair $(H_1,H_2)$, there may not exist a new pair
$(K_1,K_2)$ for the same transformation.

It is well known that the canonicity condition of a transformation
must be independent from the forms of the Hamilton functions. We
now give a theorem related with this condition. Our theorem is
three-dimensional time dependent generalization of the
two-dimensional time independent version \cite{ref:Hurley}.
\newline\newline%
{\bf Theorem 2.1.} The transformation $(\ref{imap})$ is canonoid
with respect to all Hamiltonian pairs iff
\begin{eqnarray}\label{constant}
\{X_1,X_2,X_3\}=\rm{constant}.
\end{eqnarray}
\newline%
{\bf Proof:} If we consider the fact that
\begin{eqnarray}\label{Kzero}
\epsilon_{ijk}\frac{\partial}{\partial X_i}\frac{\partial
(K_1,K_2)}{\partial (X_j,X_k)}=0,
\end{eqnarray}
it is apparent from (\ref{covariance}) that the existence of $K_1$
and $K_2$ is equivalent to
\begin{eqnarray}\label{dots}
\frac{\partial\dot{X_i}}{\partial X_i}=0.
\end{eqnarray}
Since
\begin{eqnarray}\label{dot}
\dot{X}_i(x,t)=\frac{\partial X_i}{\partial
x_j}\dot{x_j}+\frac{\partial X_i}{\partial t},
\end{eqnarray}
with the help of (\ref{NH}), (\ref{dots}) reduces to
\begin{eqnarray}\label{cond1}
\epsilon_{jkl}\frac{\partial }{\partial X_i}\left(\frac{\partial
X_i}{\partial x_j}\frac{\partial H_1}{\partial x_k}\frac{\partial
H_2}{\partial x_l}\right)+\frac{\partial }{\partial
X_i}\frac{\partial X_i}{\partial t}=0.
\end{eqnarray}
Equivalently,
\begin{eqnarray}\label{derivate}
\epsilon_{jkl}\left(\frac{\partial }{\partial X_i}\frac{\partial
X_i}{\partial x_j}\right)\frac{\partial H_1}{\partial
x_k}\frac{\partial H_2}{\partial x_l}+\epsilon_{jkl}\frac{\partial
X_i }{\partial x_j}\frac{\partial }{\partial
X_i}\left(\frac{\partial H_1}{\partial x_k}\frac{\partial
H_2}{\partial x_l}\right)+\frac{\partial }{\partial
X_i}\frac{\partial X_i}{\partial t}=0.
\end{eqnarray}
If the first transformation rule in (\ref{vfield}) is used, the
second term of (\ref{derivate}) vanishes as
\begin{eqnarray}\label{Hzero}
\epsilon_{jkl}\frac{\partial}{\partial x_j}\frac{\partial
(H_1,H_2)}{\partial (x_k,x_l)}=0.
\end{eqnarray}
If we impose the requirement that the transformation is a canonoid
transformation independent from the Hamilton functions $H_1$ and
$H_2$, the coefficients in the first term of (\ref{derivate}) must
vanish, namely
\begin{eqnarray}\label{coeff}
\frac{\partial}{\partial X_i}\frac{\partial X_i}{\partial x_j}=0.
\end{eqnarray}
The last term in (\ref{derivate}) is already Hamiltonian
independent and it gets directly zero with the condition
(\ref{coeff}). Therefore the theorem becomes equal to the
following statement
\begin{eqnarray}
\frac{\partial}{\partial X_i}\frac{\partial X_i}{\partial x_j}=0
\quad\Leftrightarrow\quad \{X_1,X_2,X_3\}=\rm{constant}.
\end{eqnarray}
It is straightforward to see, after a bit long but simple
calculation, that
\begin{eqnarray}\label{zeroB}
\partial_{X_m}\{X_1,X_2,X_3\}=0,
\end{eqnarray}
if (\ref{coeff}) is satisfied. Conversely, the explicit form of
(\ref{zeroB}), for $m=1$ for instance, is
\begin{eqnarray}\label{m1}
\partial_{X_1}\{X_1,X_2,X_3\}=\epsilon_{jkl}\frac{\partial X_2}{\partial x_k}\frac{\partial
X_3}{\partial x_l}\frac{\partial }{\partial X_i}\frac{\partial
X_i}{\partial x_j}=0.
\end{eqnarray}
Together with the other two values of $m$, (\ref{m1}) defines a
homogeneous system of linear equations for the unknowns
\begin{eqnarray}
\frac{\partial}{\partial X_i}\frac{\partial X_i}{\partial x_j}.
\end{eqnarray}
The determinant of the matrix of coefficients gives
$\{X_1,X_2,X_3\}^2$ and with the condition (\ref{constant}), the
unique solution is then the trivial one, i.e.,
(\ref{coeff}).\newline%
\null\hspace{17.5cm}$\Box$
\newline\newline%
{\bf Definition 2.2.} A \emph{canonical} transformation is a
canonoid transformation with
\begin{eqnarray}\label{necessaryandsufficient}
\{X_1,X_2,X_3\}=1.
\end{eqnarray}
Therefore a CT is a transformation preserving the fundamental
Nambu bracket
\begin{eqnarray}\label{micro}
\{x_1,x_2,x_3\}=1
\end{eqnarray}
independently from the forms of the pair $(H_1,H_2)$.
Additionally, if one employs the transformation rule
(\ref{vfield}) for (\ref{micro}), the canonicity condition gives
\begin{eqnarray}
\{x_1,x_2,x_3\}_X=1,
\end{eqnarray}
where the subscript $X$ means that the derivatives in the
expansion of the bracket are taken with respect to the new
coordinates $X_1,X_2,X_3$.

In fact, a brief definition of the CTs in the three-space is given
in Ref.~\cite{ref:Takhtajan} as a diffeomorphism of the phase
space which preserve Nambu bracket structure. But such a
definition bypasses the probability that the transformation is a
canonoid transformation.\newline\newline%
{\bf Remark 2.1.} A CT preserves the Nambu bracket of arbitrary
functions, i.e.,
\begin{eqnarray}
\{f(x,t),g(x,t),h(x,t)\}_x=\{f(x,t),g(x,t),h(x,t)\}_X.
\end{eqnarray}
According to the Remark 2.1., one gets
\begin{subequations}
\begin{eqnarray}\label{direct1}
\{X_i,H_1,H_2\}_x&=&\{X_i,H_1,H_2\}_X,\\
\{x_i,H_1,H_2\}_x&=&\{x_i,H_1,H_2\}_X.\label{direct2}
\end{eqnarray}
\end{subequations}
With the help of (\ref{vfield}), the first covariance
(\ref{direct1}) implies the first group of conditions on a CT
\begin{eqnarray}\label{firstgroup}
\frac{\partial X_i}{\partial x_l}=\frac{\partial
(x_m,x_n)}{\partial (X_j,X_k)},
\end{eqnarray}
and (\ref{direct2}) implies the second group
\begin{eqnarray}\label{secondgroup}
\frac{\partial x_i}{\partial X_l}=\frac{\partial
(X_m,X_n)}{\partial (x_j,x_k)},
\end{eqnarray}
where $(i,j,k)$ and $(l,m,n)$ are cycling indices.
(\ref{firstgroup}) and (\ref{secondgroup}) are the equations
corresponding to the so-called \emph{direct conditions} in
Hamilton formalism.

\section{Generating Functions}
We now discuss how CTs can be generated in the three-space. We
will show that to each CT corresponds a particular pair
$(F_1,F_2)$. $F_1$ and $F_2$ are the GFs of the transformation
defined on $\mathbb{R}^3\times \mathbb{R}$, and as shown in Sec.5,
they can give a complete classification of the CTs.

We start with the three-form
\begin{eqnarray}\label{3form}
\chi =dX_1\wedge dX_2\wedge dX_3.
\end{eqnarray}
When (\ref{form}) is used for every one-form in (\ref{3form}), we
get by (\ref{necessaryandsufficient}) that
\begin{eqnarray}\label{form-exp}
dX_1\wedge dX_2\wedge dX_3=dx_1\wedge dx_2\wedge
dx_3+\frac{\partial (X_{[i},X_j)}{\partial
(x_l,x_m)}\frac{\partial X_{k\,]}}{\partial t}\,dx_l\wedge
dx_m\wedge dt,
\end{eqnarray}
where the bracket [ ] stands for the cyclic sum. The substitution
of the term
\begin{eqnarray}
\frac{\partial X_i}{\partial t}=\frac{\partial (K_1,K_2)}{\partial
(X_j,X_k)}-\{X_i,H_1,H_2\}
\end{eqnarray}
obtained by (\ref{NH}), (\ref{covariance}) and (\ref{dot}), into
(\ref{form-exp}) gives ultimately that
\begin{eqnarray}\label{ultimate}
dX_1\wedge dX_2\wedge dX_3=dx_1\wedge dx_2\wedge dx_3-dH_1\wedge
dH_2\wedge dt+dK_1\wedge dK_2\wedge dt.
\end{eqnarray}
The first property that should be pointed out for (\ref{ultimate})
is that, for the time independent transformations it reduces
simply to
\begin{eqnarray}\label{condnotime}
dX_1\wedge dX_2\wedge dX_3=dx_1\wedge dx_2\wedge dx_3
\end{eqnarray}
which is an alternative test for the canonicity. Now let us
rewrite (\ref{ultimate}) as
\begin{eqnarray}
d\Omega =d(x_1dx_2\wedge dx_3-X_1dX_2\wedge dX_3-H_1dH_2\wedge
dt+K_1dK_2\wedge dt)=0.
\end{eqnarray}
We assume that the closed two-form $\Omega$ can be decomposed as
the product of two one-forms $dF_1$ and $dF_2$, then
\begin{eqnarray}\label{GFs}
dF_1\wedge dF_2=x_1dx_2\wedge dx_3-X_1dX_2\wedge
dX_3-H_1dH_2\wedge dt+K_1dK_2\wedge dt.
\end{eqnarray}
Equating the coefficients of similar basic two-forms not including
$dt$ on both sides of (\ref{GFs}) gives
\begin{eqnarray}\label{Ninterior}
&& \frac{\partial (F_1,F_2)}{\partial
(x_2,x_3)}=x_1-X_1\frac{\partial (X_2,X_3)}{\partial
(x_2,x_3)}:=A(x,t),\nonumber\\
&& \frac{\partial (F_1,F_2)}{\partial
(x_3,x_1)}=-X_1\frac{\partial
(X_2,X_3)}{\partial (x_3,x_1)}:=B(x,t),\nonumber\\
&& \frac{\partial (F_1,F_2)}{\partial
(x_1,x_2)}=-X_1\frac{\partial (X_2,X_3)}{\partial
(x_1,x_2)}:=C(x,t),
\end{eqnarray}
where the relation
\begin{eqnarray}
\frac{\partial A}{\partial x_1}+\frac{\partial B}{\partial
x_2}+\frac{\partial C}{\partial x_3}=0
\end{eqnarray}
is satisfied independently from the transformation due to the
general rule (\ref{Hzero}) written for the GFs $F_1$ and $F_2$.
(\ref{Ninterior}) is a useful set of equations in finding both GFs
and CTs: Since we have also
\begin{eqnarray}\label{ado}
\frac{\partial F_\alpha}{\partial x_{[i}}\,\frac{\partial
(F_1,F_2)}{\partial (x_j,x_{k]})}=\epsilon_{ijk}\frac{\partial
F_\alpha}{\partial x_i}\,\frac{\partial F_1}{\partial
x_j}\frac{\partial F_2}{\partial x_k}=0,
\end{eqnarray}
given CT $X_i(x)$, the GFs appear as the solution to the Pfaffian
partial differential equation
\begin{eqnarray}\label{Pfaffian2}
A(x,t)\frac{\partial F_\alpha}{\partial x_1}+B(x,t)\frac{\partial
F_\alpha}{\partial x_2}+C(x,t)\frac{\partial F_\alpha}{\partial
x_3}=0,
\end{eqnarray}
up to an additive function of $t$. Conversely, given GFs,
(\ref{Ninterior}) provides the differential equation for $X_2$ and
$X_3$
\begin{eqnarray}\label{Pfaffian3}
[A(x,t)-x_1]\frac{\partial X_\beta}{\partial
x_1}+B(x,t)\frac{\partial X_\beta}{\partial
x_2}+C(x,t)\frac{\partial X_\beta}{\partial x_3}=0\quad ,\quad
\beta =2,3.
\end{eqnarray}
Once $X_\beta (x,t)$ has been determined, the complementary part
$X_1(x,t)$ of the transformation is immediate by returning to
(\ref{Ninterior}).

The general solutions to (\ref{Pfaffian2}) and (\ref{Pfaffian3})
are arbitrary functions of some unique arguments. Hence,
$F_\alpha$ or $X_\beta$ do not specify the transformation
uniquely. However, by obeying the conventional procedure in the
textbooks, through the text we will accept these unique arguments
as the solutions so long as they are suitable for our aim.

On the other hand, in (\ref{GFs}), the coefficients of the forms
including $dt$ gives another useful relation between the GFs, the
CT and the new Hamilton functions;
\begin{eqnarray}\label{a}
 \frac{\partial (F_1,F_2)}{\partial
(x_i,t)}=-H_1\,\frac{\partial H_2}{\partial
x_i}+K_1\,\frac{\partial K_2}{\partial x_i}- X_1\,\frac{\partial
(X_2,X_3)}{\partial (x_i,t)}.
\end{eqnarray}
Given a dynamical system with $(H_1,H_2)$ and a CT, finding the
pair $(K_1,K_2)$ is another matter. In order to find the new
Hamilton functions, we consider the interior product of
$\partial_{\,t}$ and the three-form (\ref{ultimate}) resulting
\begin{eqnarray}\label{b}
\frac{\partial (K_1,K_2)}{\partial (x_i,x_j)}=\frac{\partial
(H_1,H_2)}{\partial (x_i,x_j)}+\frac{\partial
(X_{[k},X_l)}{\partial (x_i,x_j)}\,\frac{\partial X_{m]}}{\partial
t}=:f_{ij}(x,t).
\end{eqnarray}
Given $f_{ij}$, by means of (\ref{ado}) which is also valid for
the pair $(K_1,K_2)$, we obtain the differential equation
\begin{eqnarray}\label{Ksmall}
f_{\,[ij}\,\frac{\partial K_\alpha}{\partial x_{k]}}=0
\end{eqnarray}
whose solutions are the new Hamilton functions.

Alternatively, the Pfaffian partial differential equation
\begin{eqnarray}\label{Pfaffian1}
\dot{X}_i\,\frac{\partial K_\alpha}{\partial X_i}=0,
\end{eqnarray}
originated from (\ref{covariance}) and from the fact
\begin{eqnarray}\label{grule}
\frac{\partial K_\alpha}{\partial X_{[i}}\,\frac{\partial
(K_1,K_2)}{\partial (X_j,X_{k]})}=\epsilon_{ijk}\frac{\partial
K_\alpha}{\partial X_i}\,\frac{\partial K_1}{\partial
X_j}\frac{\partial K_2}{\partial X_k}=0,
\end{eqnarray}
gives the same solution pair but in terms of $X$. It is apparent
that the pairs $(F_1,F_2)$ and $(K_1,K_2)$ must also satisfy
$(\ref{a})$.

For the time independent CTs, finding the new Hamilton functions
is much easier without considering the differential equations
given above:\newline\newline%
{\bf Theorem 3.1.} If the CT is time independent, then the new
Hamiltonian pair can be found simply as
\begin{eqnarray}
(K_1(X,t),K_2(X,t))=(H_1(x(X),t),H_2(x(X),t)).
\end{eqnarray}
\newline%
{\bf Proof:}
\begin{eqnarray}\label{findingK}
\dot{X}_i=\frac{\partial X_i}{\partial
x_j}\dot{x}_j&=&\epsilon_{jmn}\frac{\partial X_i}{\partial
x_j}\frac{\partial H_1}{\partial x_m}\frac{\partial H_2}{\partial
x_n}\nonumber\\
&=&\epsilon_{jmn}\frac{\partial X_i}{\partial x_j}\frac{\partial
X_k}{\partial x_m}\frac{\partial X_l}{\partial x_n}\frac{\partial
H_1}{\partial X_k}\frac{\partial H_2}{\partial X_l}\nonumber\\
&=&\{X_i,X_k,X_l\}\frac{\partial H_1}{\partial X_k}\frac{\partial
H_2}{\partial X_l}\nonumber\\
&=&\epsilon_{ikl}\frac{\partial H_1}{\partial X_k}\frac{\partial
H_2}{\partial X_l}=\frac{\partial (H_1,H_2)}{\partial (X_k,X_l)}\nonumber\\
&=&\frac{\partial (K_1,K_2)}{\partial (X_k,X_l)},
\end{eqnarray}
where $(i,k,l)$ are cycling indices again and (\ref{vfield}) and
(\ref{explicit}) are used in the first and second lines
respectively. \newline%
\null\hspace{17.5cm}$\Box$\newline%
Note that the new Hamilton functions
$K_1$ and $K_2$ may contain $t$ explicitly due to $H_1(x,t)$ and
$H_2(x,t)$ even if the transformation is time independent.

Before concluding this section, it may be remarkable to point out
that in his original paper, as an interesting approach, Nambu
considers the CT itself as equations of motion generated by the
closed two-form
\begin{eqnarray}\label{GH}
dH(x)\wedge dG(x)=X_1(x)dx_2\wedge dx_3+X_2(x)dx_3\wedge
dx_1+X_3(x)dx_1\wedge dx_2.
\end{eqnarray}
Though (\ref{GH}) is a powerful tool to find the CT or the GFs,
its closeness property imposes the restriction
\begin{eqnarray}\label{rest}
\frac{\partial X_1}{\partial x_1}+\frac{\partial X_2}{\partial
x_2}+\frac{\partial X_3}{\partial x_3}=0
\end{eqnarray}
on the transformation. Linear CT (\ref{LCT}) satisfies the
restriction (\ref{rest}) and its analysis via (\ref{GH}) can be
found in Ref.~\cite{ref:Nambu}.
\section{Most Known Canonical Transformations and Their Generating Functions}

(i) Scaling transformation:
\begin{eqnarray}\label{scaling}
X_1=ax_1 \;,\; X_2=bx_2\;,\; X_3=cx_3\;,\;\;abc=1.
\end{eqnarray}
Since the transformation is time independent, (\ref{GFs}) becomes
\begin{eqnarray}
dF_1\wedge dF_2=0.
\end{eqnarray}
There exist three possibilities for the GFs: $F_\alpha =$
constant, $F_2=F_2(F_1)$ and $F_1=f(x),\;F_2=$ constant. We prefer
the one compatible with the usual Hamilton formalism, i.e.,
$F_\alpha =$ constant which also corresponds to the so-called
Methieu transformation \cite{ref:Whittaker}. The special case
$a=b=c=1$ is the identity transformation, of course.

As a direct application consider the Euler equations of a rigid
body \cite{ref:Nambu}
\begin{eqnarray}\label{Euler}
\dot{x_1}=x_2\frac{x_3}{I_3}-x_3\frac{x_2}{I_2},\nonumber\\
\dot{x_2}=x_3\frac{x_1}{I_1}-x_1\frac{x_3}{I_3},\nonumber\\
\dot{x_3}=x_1\frac{x_2}{I_2}-x_2\frac{x_1}{I_1},
\end{eqnarray}
where $x_i$ stands for the components of angular momentum and
$I_i$ is the moment of inertia corresponding to the related
principal axis. If we take $\gamma_i^2=-1/I_j+1/I_k$ with the
cycling indices, (\ref{Euler}) leads to
\begin{eqnarray}
\dot{x_1}=\gamma_1^2\,x_2x_3\;,\;\dot{x_2}=\gamma_2^2x_3x_1\;,
\;\dot{x_3}=\gamma_3^2\,x_1x_2,\qquad
\gamma_1^2+\gamma_2^2+\gamma_3^2=0.
\end{eqnarray}
If $\gamma_1\gamma_2\gamma_3=1$ is also satisfied, then the
equations of motion are generated by the Hamilton functions
\begin{eqnarray}
H_1=\frac{1}{2}\left(\frac{x_1^2}{\gamma_1^2}-\frac{x_2^2}{\gamma_2^2}\right)\;,\;
H_2=\frac{1}{2}\left(\frac{x_1^2}{\gamma_1^2}-\frac{x_3^2}{\gamma_3^2}\right).
\end{eqnarray}
The scaling transformation
\begin{eqnarray}
X_1=x_1/\gamma_1 \;,\; X_2=x_2/\gamma_2\;,\; X_3=x_3/\gamma_3,
\end{eqnarray}
converts the Euler system (\ref{Euler}) into the Lagrange system
\cite{ref:Chakravarty}
\begin{eqnarray}
\dot{X_1}=X_2X_3\;,\;\dot{X_2}=X_3X_1\;, \;\dot{X_3}=X_1X_2
\end{eqnarray}
which is also called Nahm's system in the theory of static
$SU(2)$-monopoles generated by the transformed Hamilton functions
\begin{eqnarray}
K_1=\frac{1}{2}\left(X_1^2-X_2^2\right)\;,\;
K_2=\frac{1}{2}\left(X_1^2-X_3^2\right).
\end{eqnarray}
\\
(ii) Linear transformations:

Three-dimensional version of the linear CT is immediate:
\begin{eqnarray}\label{LCT}
X_1&=&a_1\,x_1+a_2\,x_2+a_3\,x_3,\nonumber\\
X_2&=&b_1\,x_1+b_2\,x_2+b_3\,x_3,\nonumber\\
X_3&=&c_1\,x_1+c_2\,x_2+c_3\,x_3,
\end{eqnarray}
satisfying $a_1\,\alpha_1+a_2\,\alpha_2+a_3\,\alpha_3=1$, where
\begin{eqnarray}
\alpha_1&=&b_2\,c_3-b_3\,c_2,\nonumber\\
\alpha_2&=&b_3\,c_1-b_1\,c_3,\nonumber\\
\alpha_3&=&b_1\,c_2-b_2\,c_1.
\end{eqnarray}
The solutions to (\ref{Pfaffian2}) appear as the GFs;
\begin{eqnarray}
F_1(x)&=&\alpha_2\,x_3-\alpha_3\,x_2,\nonumber\\
F_2(x)&=&-\frac{1}{2}\,a_1\,x_1^2+
\frac{\alpha_1}{2\,\alpha_2}\,a_2\,x_2^2+
\frac{\alpha_1}{2\,\alpha_3}\,a_3\,x_3^2-a_2\,x_1\,x_2-a_3\,x_1\,x_3.
\end{eqnarray}

As an application of the linear CTs we consider the Takhtajan's
system \cite{ref:Takhtajan};
\begin{eqnarray}\label{Takhtajan}
\dot{x_1}=x_2-x_3\;,\;\dot{x_2}=x_3-x_1\;,\;\dot{x_3}=x_1-x_2.
\end{eqnarray}
The implicit solution of the system is the trajectory vector
$\textbf{r}(t)=x_1(t)\,\textbf{e}_1+x_2(t)\,\textbf{e}_2+x_3(t)\,\textbf{e}_3$
tracing out the curve which is the intersection of the sphere
$H_1=(x_1^2+x_2^2+x_3^2)/2$ and the plane $H_2=x_1+x_2+x_3$.
$\textbf{r}(t)$ makes a precession motion with a constant angular
velocity around the vector
$\textbf{N}=\textbf{e}_1+\textbf{e}_2+\textbf{e}_3$ normal to the
$H_2$ plane. The linear CT corresponding to the rotation
\begin{eqnarray}
&&X_1=\frac{1}{\sqrt{6}}\,x_1+\frac{1}{\sqrt{6}}\,x_2-\frac{2}{\sqrt{6}}\,x_3,\nonumber\\
&&X_2=-\frac{1}{\sqrt{2}}\,x_1+\frac{1}{\sqrt{2}}\,x_2,\nonumber\\
&&X_3=\frac{1}{\sqrt{3}}\,x_1+\frac{1}{\sqrt{3}}\,x_2+\frac{1}{\sqrt{3}}\,x_3
\end{eqnarray}
coincides $\textbf{N}$ with the $\textbf{e}_3$ axis. The new
system is then given by the well-known equations of motion of the
Harmonic oscillator
\begin{eqnarray}
\dot{X_1}=\sqrt{3}\,X_2\;,\;\dot{X_2}=-\sqrt{3}\,X_1\;,\;\dot{X_3}=0
\end{eqnarray}
with the Hamilton functions $K_1=(X_1^2+X_2^2+X_3^2)/2$ and
$K_2=\sqrt{3}\,X_3$. Therefore inverse of the transformation
provides directly an explicit solution to the original system.\\

(iii) Gauge transformations:

We will define the gauge transformation in our three-dimensional
phase space as a model transformation which is similar to the case
in the usual Hamilton formalism:
\begin{eqnarray}\label{gauge1}
X_1=x_1\;,\;X_2=x_2+f_1(x_1)\;,\;X_3=x_3+f_2(x_1),
\end{eqnarray}
where $f_1(x_1)$  and $f_2(x_1)$ are arbitrary functions
determined by the GF. Since
\begin{eqnarray}
A(x)=0\;,\;B(x)=x_1\,\frac{\partial f_1}{\partial
x_1}\;,\;C(x)=x_1\,\frac{\partial f_2}{\partial x_1},
\end{eqnarray}
(\ref{Pfaffian2}) provides us the GFs as the following form
\begin{eqnarray}
F_1(x)=x_2\,\frac{\partial f_2}{\partial x_1}-x_3\,\frac{\partial
f_1}{\partial x_1}\;,\;F_2(x)=-\frac{1}{2}\,x_1^2.
\end{eqnarray}
By keeping ourselves in this argument, other possible gauge
transformation types can be constructed easily. For instance, a
second kind of gauge transformation can be defined by
\begin{eqnarray}\label{gauge2}
X_1=x_1+g_1(x_2)\;,\;X_2=x_2\;,\;X_3=x_3+g_2(x_2)
\end{eqnarray}
and it is generated by $F_1=g_1(x_2)\,x_3$ and $F_2=x_2$. Another
type is
\begin{eqnarray}\label{gauge3}
X_1=x_1+h_1(x_3)\;,\;X_2=x_2+h_2(x_3)\;,\;X_3=x_3
\end{eqnarray}
and it is generated by $F_1=h_1(x_3)\,x_2$ and $F_2=-x_3$.
\\

(iv) Point transformations:

Our model transformation which is similar to the Hamilton
formalism again will be in the form
\begin{eqnarray}\label{point1}
X_1=f_1(x_1)\;,\;X_2=f_2(x_1)\,x_2\;,\;X_3=f_3(x_1)\,x_3,
\end{eqnarray}
where $f_1$, $f_2$ and $f_3$ are arbitrary functions satisfying
\begin{eqnarray}
\frac{\partial f_1}{\partial x_1}\,f_2\,f_3=1.
\end{eqnarray}
(\ref{Ninterior}) says that
\begin{eqnarray}
A(x)=x_1-f_1\,f_2\,f_3\;,\;B(x)=x_2\,f_1\,f_3\,\frac{\partial
f_2}{\partial x_1}\;,\;C(x)=x_3\,f_1\,f_2\,\frac{\partial
f_3}{\partial x_1},
\end{eqnarray}
and to find the GFs we use (\ref{Pfaffian2}) of course, hence
\begin{eqnarray}
F_1(x)=x_2\,\exp \left(-\int
\frac{B}{C\,x_2}dx_1\right)\;,\;F_2(x)=x_3\,\exp \left(-\int
\frac{A}{C\,x_3}dx_1\right),
\end{eqnarray}
where
\begin{eqnarray}
\exp \left[-\int
\frac{1}{C}\,\left(\frac{A}{x_3}+\frac{B}{x_2}\right)dx_1\right]=C.
\end{eqnarray}
Other possible types of the point transformation;
\begin{eqnarray}
X_1=g_1(x_2)\,x_1\;,\;X_2=g_2(x_2)\;,\;X_3=g_3(x_2)\,x_3,
\end{eqnarray}
and
\begin{eqnarray}
X_1=h_1(x_3)\,x_1\;,\;X_2=h_2(x_3)\,x_2\;,\;X_3=h_3(x_3)
\end{eqnarray}
give surprisingly constant GFs.
\\

(v) Rotation in $\mathbb{R}^3$:

This last example is chosen as time dependent so that it makes the
procedure through a CT more clear. Consider again the system
(\ref{Takhtajan}) together with the CT
\begin{eqnarray}
X_1=x_1\;,\;X_2=x_2\,\cos t+x_3\,\sin t\;,\;X_3=-x_2\,\sin
t+x_3\,\cos t
\end{eqnarray}
corresponding to the rotation about the $x_1$ axis. The first
attempt to determine the GFs is to consider (\ref{Pfaffian2}).
Since $A(x)=0$, $B(x)=0$ and $C(x)=0$, that equation does not give
enough information on the pair $(F_1,F_2)$. Still things can be
put right by considering first (\ref{Ksmall}). For our case it
yields
\begin{eqnarray}
(x_2-x_3)\,\frac{\partial K_\alpha}{\partial
x_1}+(2x_3-x_1)\,\frac{\partial K_\alpha}{\partial x_2}+
(x_1-2x_2)\,\frac{\partial K_\alpha}{\partial x_3}=0
\end{eqnarray}
with the solution
\begin{eqnarray}
K_1=\frac{1}{2}(x_1^2+x_2^2+x_3^2)\;,\;K_2=2x_1+x_2+x_3.
\end{eqnarray}
Note that one gets, with the aid of the inverse transformation,
that
\begin{eqnarray}
K_1=\frac{1}{2}(X_1^2+X_2^2+X_3^2)\;,\;K_2=2X_1+(\cos t+\sin
t)X_2+(\cos t-\sin t)X_3
\end{eqnarray}
and this is also the solution to (\ref{Pfaffian1}). Now the right
hand side of (\ref{a}) is explicit and the solution
\begin{eqnarray}
F_1=\frac{x_1}{2}\left(
\frac{x_1^2}{3}+x_2^2+x_3^2\right)\;,\;F_2=t
\end{eqnarray}
also satisfies (\ref{Ninterior}) or (\ref{Pfaffian2}).
\section{Generating Functions of Type}
A CT may admit various independent triplets on
$\mathbb{R}^3\times\mathbb{R}$ apart from $(x_1,x_2,x_3)$ or
$(X_1,X_2,X_3)$. Two main groups are possible; first one is
$(x_i,x_j,X_k)$, and the second one is $(X_i,X_j,x_k)$, where
$i\neq j$ and every group contains obviously nine triplets. In
order to show how one can determine the transformation types, two
different types of them are treated explicitly. The calculation
scheme is the same for all possible types which is listed in Table
~\ref{table1}.

\begin{table}[ph]
\caption{Types of the canonical transformations in six kinds.
($r=1,...,6$ and $U=H_1dH_2\wedge dt-K_1dK_2\wedge dt$).}

{\begin{tabular}{@{}cc@{}} %\toprule
Independent variables & $(dF_1\wedge dF_2)r$\\
\hline %
$x_1,x_2,X_1$ & \\
$x_1,x_2,X_2$ & $df_1\wedge df_2+d(x_1x_3)\wedge dx_2 =
x_3dx_1\wedge
dx_2-X_1dX_2\wedge dX_3-U$\\
$x_1,x_2,X_3$ & \\
\hline %
$x_1,x_3,X_1$ & \\
$x_1,x_3,X_2$ & $df_1\wedge df_2-d(x_1x_2)\wedge dx_3 =
x_2dx_3\wedge
dx_1-X_1dX_2\wedge dX_3-U$\\
$x_1,x_3,X_3$ & \\
\hline %
$x_2,x_3,X_1$ & \\
$x_2,x_3,X_2$ & $df_1\wedge df_2= x_1dx_2\wedge
dx_3-X_1dX_2\wedge dX_3-U$\\
$x_2,x_3,X_3$ & \\
\hline %
$X_1,X_2,x_1$ & \\
$X_1,X_2,x_2$ & $df_1\wedge df_2-d(X_1X_3)\wedge dX_2 =
x_1dx_2\wedge
dx_3-X_3dX_1\wedge dX_2-U$\\
$X_1,X_2,x_3$ & \\
\hline %
$X_1,X_3,x_1$ & \\
$X_1,X_3,x_2$ & $df_1\wedge df_2+d(X_1X_2)\wedge dX_3 =
x_1dx_2\wedge
dx_3-X_2dX_3\wedge dX_1-U$\\
$X_1,X_3,x_3$ & \\
\hline %
$X_2,X_3,x_1$ & \\
$X_2,X_3,x_2$ & $df_1\wedge df_2= x_1dx_2\wedge
dx_3-X_1dX_2\wedge dX_3-U$\\
$X_2,X_3,x_3$ & \\
\hline %
\end{tabular}\label{table1}}
\end{table}

First, we consider the triplet $(x_1,x_2,X_3)$. Then if every term
in (\ref{GFs}) is written in terms of $(x_1,x_2,X_3)$, the
equivalence of related coefficients of the components on both
sides of that equation amounts to
\begin{eqnarray}\label{typexyZ}
\frac{\partial (f_1,f_2)}{\partial
(x_1,x_2)}&=&-x_1\,\frac{\partial x_3}{\partial x_1},\nonumber\\
\frac{\partial (f_1,f_2)}{\partial
(X_3,x_1)}&=&X_1\,\frac{\partial X_2}{\partial x_1},\nonumber\\
\frac{\partial (f_1,f_2)}{\partial
(x_2,X_3)}&=&x_1\,\frac{\partial x_3}{\partial
X_3}-X_1\,\frac{\partial X_2}{\partial x_2},
\end{eqnarray}
and
\begin{eqnarray}\label{typexyZt}
\frac{\partial (f_1,f_2)}{\partial (x_1,t)}&=&-H_1\,\frac{\partial
H_2}{\partial x_1}+
K_1\,\frac{\partial K_2}{\partial x_1},\nonumber\\
\frac{\partial (f_1,f_2)}{\partial (x_2,t)}&=&-H_1\,\frac{\partial
H_2}{\partial x_2}+
K_1\,\frac{\partial K_2}{\partial x_2}+x_1\,\frac{\partial x_3}{\partial t},\nonumber\\
\frac{\partial (f_1,f_2)}{\partial (X_3,t)}&=&-H_1\,\frac{\partial
H_2}{\partial X_3}+ K_1\,\frac{\partial K_2}{\partial
X_3}+X_1\,\frac{\partial X_2}{\partial t},
\end{eqnarray}
where $f_\alpha =F_\alpha (x_1,x_2,x_3(x_1,x_2,X_3,t),t)$. Given
GFs $f_1$ and $f_2$, these equations do not give always complete
information on the transformation. But consider the rearrangement
of (\ref{typexyZ})
\begin{eqnarray}\label{hacivat}
\frac{\partial (f_1,f_2)}{\partial (x_1,x_2)}+\frac{\partial (x_1
x_3,x_2)}{\partial
(x_1,x_2)}&=&x_3,\nonumber\\
\frac{\partial (f_1,f_2)}{\partial (X_3,x_1)}+\frac{\partial (x_1
x_3,x_2)}{\partial
(X_3,x_1)}&=&X_1\,\frac{\partial X_2}{\partial x_1},\nonumber\\
\frac{\partial (f_1,f_2)}{\partial (x_2,X_3)}+\frac{\partial (x_1
x_3,x_2)}{\partial (x_2,X_3)}&=&-X_1\,\frac{\partial X_2}{\partial
x_2},
\end{eqnarray}
which is equivalent to
\begin{eqnarray}\label{f12}
df_1\wedge df_2+d(x_1x_3)\wedge dx_2=&&x_3 dx_1\wedge
dx_2-X_1dX_2\wedge dX_3\nonumber\\
&&-H_1\,dH_2\wedge dt+K_1dK_2\wedge dt.
\end{eqnarray}
For the functions $F_\alpha (x_1,x_2,X_3,t)$ which are the
solutions to the differential equation
\begin{eqnarray}\label{GFtype}
X_1\,\frac{\partial X_2}{\partial x_2}\frac{\partial
F_\alpha}{\partial x_1}-X_1\,\frac{\partial X_2}{\partial
x_1}\frac{\partial F_\alpha}{\partial x_2}-x_3\frac{\partial
F_\alpha}{\partial X_3}=0
\end{eqnarray}
obtained from (\ref{hacivat}); (\ref{f12}) leads to
\begin{eqnarray}\label{F12}
(dF_1\wedge dF_2)_1=x_3 dx_1\wedge dx_2-X_1dX_2\wedge
dX_3-H_1\,dH_2\wedge dt+K_1dK_2\wedge dt
\end{eqnarray}
corresponding to the our first kind transformation. Note, as can
be seen from Table~\ref{table1}, that the first kind contains
three types. Now $x_3$ is immediate by
\begin{eqnarray}
\frac{\partial (F_1,F_2)}{\partial (x_1,x_2)}=x_3,
\end{eqnarray}
and for $X_2$ one needs to solve
\begin{eqnarray}\label{X2}
\left[ \frac{\partial (F_1,F_2)}{\partial
(x_2,X_3)}\right]\frac{\partial X_2}{\partial x_1}-\left[
\frac{\partial (F_1,F_2)}{\partial (X_3,x_1)}
\right]\frac{\partial X_2}{\partial x_2}=0
\end{eqnarray}
which is originated from (\ref{hacivat}) again. Note that the
equivalence of (\ref{f12}) and (\ref{F12}) does not imply in
general $F_1=f_1+x_1x_3$ and $F_2=f_2+x_2$ unless $df_1\wedge
dx_2=df_2\wedge d(x_1x_3)$. On the other hand, for the
transformations $f_2=x_2$, the equivalence
\begin{eqnarray}\label{simplest}
dF_1\wedge dF_2=d(f_1+x_1x_3)\wedge dx_2
\end{eqnarray}
is always possible. To be more explicit about this remark,
consider the CT
\begin{eqnarray}
X_1=x_1+x_2 \;,\; X_2=x_2+x_3 \;,\; X_3=x_3.
\end{eqnarray}
If the general solutions of (\ref{Pfaffian2}) are taken as the
independent functions $F_1=x_2x_3\,$,$\,F_2=x_2$, then the
corresponding functions of type become $f_1=x_2X_3\,$,$\,f_2=x_2$.
Hence by the virtue of (\ref{simplest}) the GFs are
\begin{eqnarray}\label{gfs}
F_1=(x_1+x_2)X_3\;,\;F_2=x_2.
\end{eqnarray}
Conversely, (\ref{gfs}) generates, via (\ref{hacivat}) and
(\ref{X2}), the CT
\begin{eqnarray}
X_1=x_1+x_2 \;,\; X_2=x_2+h(x_3) \;,\; X_3=x_3.
\end{eqnarray}

Second, consider the triplet $(x_2,x_3,X_1)$. This time, for
$f_\alpha (x_2,x_3,X_1,t)$, (\ref{GFs}) says
\begin{eqnarray}
\frac{\partial (f_1,f_2)}{\partial
(x_2,x_3)}&=&x_1-X_1\,\frac{\partial (X_2 , X_3)}{\partial
(x_2,x_3)},\nonumber\\
\frac{\partial (f_1,f_2)}{\partial
(X_1,x_2)}&=&-X_1\,\frac{\partial (X_2 , X_3)}{\partial
(X_1,x_2)},\nonumber\\
\frac{\partial (f_1,f_2)}{\partial
(x_3,X_1)}&=&-X_1\,\frac{\partial (X_2 , X_3)}{\partial (x_3,X_1)}
\end{eqnarray}
similar to (\ref{Ninterior}) and
\begin{eqnarray}
 \frac{\partial (f_1,f_2)}{\partial
(\xi,t)}=-H_1\,\frac{\partial H_2}{\partial
\xi}+K_1\,\frac{\partial K_2}{\partial \xi}- X_1\,\frac{\partial
(X_2,X_3)}{\partial (\xi,t)},\qquad \xi =x_2,x_3,X_1,
\end{eqnarray}
similar to (\ref{a}). This last system of equations says that
\begin{eqnarray}
df_1\wedge df_2&=&
dF_1(x_2,x_3,X_1,t)\wedge dF_2(x_2,x_3,X_1,t)\nonumber\\
&=&x_1 dx_2\wedge dx_3-X_1\,dX_2\wedge dX_3-H_1dH_2\wedge
dt+K_1dK_2\wedge dt.
\end{eqnarray}
and therefore
\begin{eqnarray}
f_\alpha (x_2,x_3,X_1,t)=F_\alpha (x_2,x_3,X_1,t).
\end{eqnarray}
Note that $F_\alpha (x_2,x_3,X_1,t)$ serves just like the GF of
first type $F_1(q,Q,t)$ of the usual Hamilton formalism. As can be
seen in the Table 1, there are six GFs of this type. The example
given above obeys also this type of transformation.

As a further consequence, one should note that a CT may be of
different types at the same time. For example the scaling
transformation given in Sec.4 admits four types simultaneously:
\begin{eqnarray}
&&F_1=\frac{1}{c}\,x_1X_3,\;F_2=x_2,\nonumber\\
&&F_1=-\frac{1}{b}\,x_1X_2,\;F_2=x_3,\nonumber\\
&&F_1=-c\,x_3X_1,\;F_2=X_2,\nonumber\\
&&F_1=b\,x_2X_1,\;F_2=X_3.
\end{eqnarray}

\section{Infinitesimal Canonical Transformations}
In the two-dimensional phase space of the usual Hamilton
formalism, ICTs are given by the variations in the first order
\begin{eqnarray}
Q &=& q+\epsilon\eta_1(q,p)=q+\epsilon \{ q,G\} = q+\epsilon
\frac{\partial G}{\partial
p},\nonumber\\
P &=&p+\epsilon\eta_2(q,p)=p+\epsilon \{ p,G\}= p-\epsilon
\frac{\partial G}{\partial q},
\end{eqnarray}
where $\epsilon$ is a continuous parameter and $G(q,p)$ is the GF
of the ICT. The canonicity condition implies
\begin{eqnarray}\label{conda}
\frac{\partial \eta_1}{\partial q}+\frac{\partial \eta_2}{\partial
p}=0
\end{eqnarray}
up to the first order of $\epsilon$. Following the same practice,
these results can be extended to the three-space. An ICT in the
three-dimensional phase space would then be proposed as
\begin{eqnarray}
X_i=x_i+\epsilon \,f_i(x)=x_i+\epsilon
\{x_i,G_1,G_2\}=x_i+\epsilon \frac{\partial (G_1,G_2)}{\partial
(x_j,x_k)},
\end{eqnarray}
where $G_1(x)$ and $G_2(x)$ generate directly the ICT via
\begin{eqnarray}
dG_1\wedge dG_2=f_1\,dx_2\wedge dx_3+f_2\,dx_3\wedge
dx_1+f_3\,dx_1\wedge dx_2.
\end{eqnarray}
One can check easily that, similar to (\ref{conda}), the
canonicity condition (\ref{condnotime}) implies
\begin{eqnarray}
\frac{\partial f_1(x)}{\partial x_1}+\frac{\partial
f_2(x)}{\partial x_2}+\frac{\partial f_3(x)}{\partial x_3}=0
\end{eqnarray}
up to the first order of $\epsilon$ again.

It is well known that an ICT is a transformation depending on a
parameter that moves the system infinitesimally along a trajectory
in phase space and therefore a finite CT is the sum of an infinite
succession of ICTs giving by the well known expansion
\begin{eqnarray}
\phi =\varphi +\epsilon \{\varphi
,G\}+\frac{\epsilon^2}{2!}\{\{\varphi ,G\},G\}
+\frac{\epsilon^3}{3!}\{\{\{\varphi ,G\},G\},G\}+\cdots
\end{eqnarray}
where $\phi =Q,P$ and $\varphi =q,p$ in turn. With the same
arguments used for the two-dimensional phase space, the
transformation equation of a finite CT generated by the GFs $G_1$
and $G_2$ will correspond to
\begin{eqnarray}\label{exp}
X_i&=&x_i+\epsilon
\{x_i,G_1,G_2\}+\frac{\epsilon^2}{2!}\{\{x_i,G_1,G_2\},G_1,G_2\}\nonumber\\
&&+\frac{\epsilon^3}{3!}\{\{\{x_i,G_1,G_2\},G_1,G_2\},G_1,G_2\}+\cdots
.
\end{eqnarray}
Equivalently, if we define the vector field
\begin{eqnarray}
\hat{V}_G=f_1(x)\,\partial_{x_1}+f_2(x)\,\partial_{x_2}+f_3(x)\,\partial_{x_3},
\end{eqnarray}
it is easy to see that the same transformation is given by
\begin{eqnarray}\label{ICT}
e^{\epsilon\,\hat{V}_G} x_i=X_i.
\end{eqnarray}

We can give a specific example showing that this construction
actually works. For this aim we consider the CT
\begin{eqnarray}
X_1=x_1,\quad X_2=x_2+\epsilon x_3,\quad X_3=x_3-\epsilon x_2.
\end{eqnarray}
The transformation is generated by GFs
\begin{eqnarray}
G_1(x)=\frac{1}{2}(x_2^2+x_3^2),\quad G_2(x)=x_1
\end{eqnarray}
or by vector field
\begin{eqnarray}
\hat{V}_G=x_3\,\partial_{x_2}-x_2\,\partial_{x_3}
\end{eqnarray}
which is the generator of rotation about $x_1$ axis. Therefore it
is immediate by means of (\ref{exp}) or (\ref{ICT}) that our
finite CT is
\begin{eqnarray}
X_1=x_1,\quad X_2=x_2\cos\epsilon+x_3\sin\epsilon,\quad
X_3=-x_2\sin\epsilon+x_3\cos\epsilon ,
\end{eqnarray}
where the parameter $\epsilon$ stands clearly for the rotation
angle.
\section{Decomposition of the Transformations}

In classical mechanics a conjecture states surprisingly that any
CT in a two dimensional phase space can be decomposed into some
sequence of two principal CTs \cite{ref:Leyvraz}. These are linear
and point CTs. Proceeding elaborations of this conjecture in
quantum mechanics led to a triplet as a wider class including
gauge, point and interchanging transformations
\cite{ref:Deenen,ref:Anderson}. One can check that the same
triplet can also be used for the classical CTs. Without giving so
many examples here, we give a particular one for the sake of
motivation: Consider the CT
\begin{eqnarray}
q\rightarrow p^2-\frac{q^2}{4p^2},\qquad p\rightarrow
-\frac{q}{2p}
\end{eqnarray}
converting the system with linear potential $H_0=p^2+q$ into the
free particle $H_1=p^2$. (In this section, we prefer using the map
representation of CTs so that we can perform easily the
transformation steps). The decomposition of the transformation can
be achieved by the following five steps in turn;
\begin{eqnarray}
&&{\rm 1.\;interchange}\qquad q\rightarrow p,\quad p\rightarrow
-q,\nonumber\\
&&{\rm 2.\;gauge}\qquad q\rightarrow q,\quad p\rightarrow
p-q^2,\nonumber\\
&&{\rm 3.\;interchange}\qquad q\rightarrow -p,\quad p\rightarrow
q,\nonumber\\
&&{\rm 4.\;point}\qquad q\rightarrow q^2,\quad p\rightarrow
p/(2q),\nonumber\\
&&{\rm 5.\;interchange}\qquad q\rightarrow -p,\quad p\rightarrow q
\end{eqnarray}
corresponding symbolically to the sequence from right to left
\begin{eqnarray}
{\cal S}={\cal I}_3\,{\cal P}\,{\cal I}_2\,{\cal G}\,{\cal I}_1.
\end{eqnarray}
As a challenging problem, the statement has not been proven in a
generic framework yet. But even though it is not true for every
CT, it applies to a huge number of CTs. Parallel to the
presentation, we will show that the discussion also applies to the
CTs in the three-space.

First we will decompose the linear CT (\ref{LCT}). Before doing
this note that all the three types (\ref{gauge1}), (\ref{gauge2}),
(\ref{gauge3}) of gauge transformation can be generated by the GFs
\begin{eqnarray}
\hat{V}_{G_1}&=&f_1(x_1)\partial_{x_2}+f_2(x_1)\partial_{x_3},\nonumber\\
\hat{V}_{G_2}&=&g_1(x_2)\partial_{x_1}+g_2(x_2)\partial_{x_3},\nonumber\\
\hat{V}_{G_3}&=&h_1(x_3)\partial_{x_1}+h_2(x_3)\partial_{x_2}
\end{eqnarray}
respectively when considering (\ref{ICT}). Now for the choices
\begin{eqnarray}
&&f_1(x_1)=\lambda_1\,x_1\;,\;f_2(x_1)=\lambda_2\,x_1,\nonumber\\
&&g_1(x_2)=\mu_1\,x_2\;,\;g_2(x_2)=\mu_2\,x_2,\nonumber\\
&&h_1(x_3)=\nu_1\,x_3\;,\;h_2(x_3)=\nu_2\,x_3,
\end{eqnarray}
the sequence
\begin{eqnarray}\label{decomplinear}
{\cal S}_L={\cal P}\,{\cal G}_3\,{\cal G}_2\,{\cal G}_1\,
\end{eqnarray}
where ${\cal P}$ stands for the point transformation generating
the scaling transformation (\ref{scaling}), generates in turn the
transformation chain
\begin{eqnarray}
&&{\rm 1.\;gauge}\qquad x_1\rightarrow x_1,\quad x_2\rightarrow
x_2+\lambda_1x_1,\quad x_3\rightarrow x_3+\lambda_2x_1,\nonumber\\
&&{\rm 2.\;gauge}\qquad x_1\rightarrow x_1+\mu_1x_2,\quad
x_2\rightarrow
x_2,\quad x_3\rightarrow x_3+\mu_2x_2,\nonumber\\
&&{\rm 3.\;gauge}\qquad x_1\rightarrow x_1+\nu_1x_3,\quad
x_2\rightarrow
x_2+\nu_2x_3,\quad x_3\rightarrow x_3,\nonumber\\
&&{\rm 4.\;point}\qquad x_1\rightarrow ax_1,\quad x_2\rightarrow
bx_2,\quad x_3\rightarrow cx_3.
\end{eqnarray}
Application of (\ref{decomplinear}) to the coordinates $(x_1, x_2,
x_3)$ gives thus the linear CT
\begin{eqnarray}
X_1&=&a\,x_1+b\,\mu_1\,x_2+c\,(\nu_1+\mu_1\,\nu_2)x_3,\nonumber\\
X_2&=&a\,\lambda_1\,x_1+b\,(1+\lambda_1\,\mu_1)\,x_2+
c\,[\lambda_1\,\nu_1+(1+\lambda_1\,\mu_1)\,\nu_2]\,x_3,\nonumber\\
X_3&=&a\,\lambda_2\,x_1+b\,(\mu_2+\mu_1\,\lambda_2)\,x_2+
c\,[1+\lambda_2\,\nu_1+(\mu_2+\mu_1\,\lambda_2)\,\nu_2]\,x_3.
\end{eqnarray}

The next example is related with the cylindrical coordinate
transformation
\begin{eqnarray}\label{c}
X_1=\frac{1}{2}(x_1^2+x_2^2)\;,\;X_2=\tan^{-1}\frac{x_2}{x_1}\;,\;X_3=x_3.
\end{eqnarray}
The sequence
\begin{eqnarray}
&&{\rm 1.\;interchange}\qquad x_1\rightarrow -x_2,\quad
x_2\rightarrow
x_1,\quad x_3\rightarrow x_3,\nonumber\\
&&{\rm 2.\;point}\qquad x_1\rightarrow \tan^{-1}x_1,\quad
x_2\rightarrow
(1+x_1^2)x_2,\quad x_3\rightarrow x_3,\nonumber\\
&&{\rm 3.\;interchange}\qquad x_1\rightarrow x_2,\quad
x_2\rightarrow
-x_1,\quad x_3\rightarrow x_3,\nonumber\\
&&{\rm 4.\;point}\qquad x_1\rightarrow x_1^2/2,\quad
x_2\rightarrow x_2/x_1,\quad x_3\rightarrow x_3,
\end{eqnarray}
which can be written in the compact form
\begin{eqnarray}
{\cal S}_C={\cal P}_2\,{\cal I}_2\,{\cal P}_1\,{\cal I}_1
\end{eqnarray}
is the decomposition of (\ref{c}).

\section*{Acknowledgements}

This work was supported by T\"{U}B\.{I}TAK (Scientific and
Technical Research Council of Turkey) under contract 107T370.

\end{document}